\documentclass[graybox]{svmult}

\usepackage{type1cm}        % activate if the above 3 fonts are
\usepackage{makeidx}         % allows index generation
\usepackage{graphicx}        % standard LaTeX graphics tool
\usepackage{multicol}        % used for the two-column index
\usepackage[bottom]{footmisc}% places footnotes at page bottom

\usepackage{newtxtext}       % 
\usepackage[varvw]{newtxmath}       % selects Times Roman as basic font

\DeclareMathOperator{\Ker}{Ker}
\DeclareMathOperator{\Attr}{Attr}
\DeclareMathOperator\supp{supp}
\DeclareMathOperator\Fix{Fix}
\DeclareMathOperator{\Ran}{Ran}
\DeclareMathOperator{\spec}{spect}

\usepackage{physics}
\usepackage{hyperref}
\usepackage{url}
\usepackage[nocompress]{cite}
\makeindex            

\begin{document}

\title*{Asymptotic Dynamics of Open Quantum Systems and Modular Theory}

\author{Daniele Amato, Paolo Facchi and Arturo Konderak}

\institute{D. Amato, P. Facchi and A. Konderak \at Dipartimento di Fisica, Universit\`a di Bari, I-70126 Bari, Italy\\and Istituto Nazionale di Fisica Nucleare, Sezione di Bari, I-70126 Bari, Italy\\
D. Amato\\
\email{daniele.amato@ba.infn.it}
\\ P. Facchi\\ \email{paolo.facchi@ba.infn.it}
\\
A. Konderak\\ \email{arturo.konderak@ba.infn.it}}

\maketitle

\abstract{In this Article, several aspects of the asymptotic dynamics of finite-dimensional open quantum systems are explored. First, after recalling a structure theorem for the peripheral map, we discuss sufficient conditions and a characterization for its unitarity. Interestingly, this is not always guaranteed due to the presence of permutations in the structure of the asymptotic map. Then, we show the connection between the asymptotic map and the modular theory by Tomita and Takesaki. }

\begin{section}{Motivation}
	The study of the dynamics of open quantum systems has received a strong boost by the recent developments in quantum technologies~\cite{wei2018efficient}. In particular, understanding and controlling the decoherence effects, arising from  nontrivial interactions of the system under interest with its surroundings, are needed for achieving optimal performance in quantum computers~\cite{zurek2003decoherence}.

	Moreover, the asymptotic evolution of the system plays a crucial role in reservoir engineering~\cite{Zoller_res_eng, Wolf_res_eng}, e.g.\ in the preparation of a target quantum state by relaxation of a system properly coupled to the bath. In this context, as well as for information protection and processing tasks~\cite{zanardi1997noiseless,zanardi2014coherent},  a one-dimensional attractor subspace, namely a unique steady state towards which the dynamics converges, is of little use for practical purposes. This has led many studies to focus on the asymptotic discrete-time evolution of open systems in a general setting, both for finite-dimensional~\cite{jex_st_2012,jex_st_2018,wolf2010inverse,wolf2012quantum,baumgartner2012structures,ticozzi_inv,albert2019asymptotics} and infinite-dimensional systems~\cite{carbone2016irreducible,carbone2020period,girotti2022absorption}. 
	
	Also, the non-unitary continuous-time dynamics at large times was investigated in the literature of the last twenty years, see e.g.~\cite{baumgartner2008analysis1,baumgartner2008analysis2,albert2014symmetries,fagnola_2001,agredo2014decoherence}, with a particular attention to quantum dynamical semigroups~\cite{Spohn_77}, completely characterized in the two seminal papers~\cite{GKS_76,Lindblad_76} from 1976.  However, the strong interest in the study of the stationary states of quantum dynamical semigroups goes back to the seventies~\cite{Spohn_77,Frigerio_78,Frigerio_Verri_82,Spohn_rev}, mainly motivated by the problem of irreversibility in quantum statistical mechanics~\cite{ingarden1975connection,evans1977dilations}.  
	
	This Article is organized as follows. After recalling some basic concepts about quantum channels (Subsection~\ref{QC}) and Tomita-Takesaki modular theory (Subsection~\ref{tomita_intro}), we will discuss a structure theorem (Theorem~\ref{Wolf_struc}) by Perez-Garcia and Wolf~\cite{wolf2010inverse,wolf2012quantum} for the attractor subspace and the action of the channel on this subspace. Then, in Section~\ref{unit_per}, we will present sufficient conditions (Theorem~\ref{suff_unit}) and a characterization (Theorem~\ref{ch_unit_as_dyn}) under which the asymptotic dynamics is unitary. Finally, in Section~\ref{Wolf-TT}, we will study the connection between the structure theorem and modular theory, recently already explored  in the continuous-time setting by Longo~\cite{longo2020emergence}.
	
	\end{section}
\begin{section}{Preliminaries}
	\label{asympt}
	\begin{subsection}{Quantum channels}
		\label{QC}
		
		In this Section we  set up the notation and recall some known concepts about the dynamics of finite-dimensional open quantum systems. 
		
		The state of an open quantum system is given by a density operator $\rho$, i.e.\ a positive semidefinite operator of unit trace on $\mathcal{H}$, the system Hilbert space with dimension $d$. In the discrete-time point of view, usually adopted in quantum information theory~\cite{nielsen2002quantum}, the evolution in the Schr\"odinger picture of an open quantum system in the unit time is described by a \emph{quantum channel} $\Phi$, namely a \emph{completely positive trace-preserving} map on $\mathcal{B}(\mathcal{H})$~\cite{heinosaari2011mathematical}. We will denote with $\mathcal{B}(\mathcal{H},\mathcal{K})$ the space of bounded operators from the Hilbert space $\mathcal{H}$ to the Hilbert space $\mathcal{K}$ and, in particular, $\mathcal{B}(\mathcal{H}) = \mathcal{B}(\mathcal{H}, \mathcal{H})$.
		
		The adjoint map $\Phi^{\dagger}$ of the channel $\Phi$ is defined with respect to  the Hilbert-Schmidt scalar product $\braket{\cdot}{\cdot}_{\mathrm {HS}}$  as
		\begin{equation}
			\braket{A}{\Phi(B)}_{\mathrm {HS}}=\langle\Phi^\dagger(A) |B\rangle_{\mathrm {HS}},\quad A,B \in \mathcal{B}(\mathcal{H}).
		\end{equation}
		$\Phi^{\dagger}$ is a completely positive unital map on $\mathcal{B}(\mathcal{H})$ and describes the system dynamics in the Heisenberg picture, i.e.\ the evolution of the system observables.  
		
		Coming back to the Schr\"odinger picture, if the system is prepared in the initial state $\rho(0)$, the evolved state $\rho(n)$ at time $t=n\in \mathbb{N}$ will be given by $\rho(n)=\Phi^{n}(\rho(0))$. Consequently the system dynamics is described by a sequence of states $\rho(0),\rho(1),\dots ,\rho(n)$, called a quantum Markov chain~\cite{jex_st_2012}.

		The spectrum $\spec(\Phi)$ of a quantum channel $\Phi$ has the following three major features~\cite{wolf2012quantum}
		\begin{itemize}
			\item $1 \in \spec(\Phi)$,
			\item $\lambda \in \spec(\Phi) \Rightarrow \bar{\lambda}\in \spec(\Phi)$,
			\item $\spec(\Phi)\subseteq \{ \lambda \in \mathbb{C} \,| \, |\lambda| \leq 1 \}$.
		\end{itemize}
		
		The asymptotic dynamics is obtained in the large $n$-limit and it ends inside the asymptotic, peripheral
		or \emph{attractor subspace} of $\Phi$, defined as
		\begin{equation}
			\Attr(\Phi)= \bigoplus_{\lambda \in \spec_P(\Phi)} \Ker(\Phi - \lambda\mathsf{1}),
		\end{equation}
			i.e.\ the direct sum of the eigenspaces belonging to the \emph{peripheral eigenvalues}
		\begin{equation}
			\lambda \in \spec_{P}(\Phi)= \{ \lambda \in \spec(\Phi)  \,|\, |\lambda|=1 \}.
		\end{equation}
		In particular, if $\lambda$ is a primitive $M$-th root of unity, 
then the corresponding eigenoperator satisfies
		\begin{equation}
		\Phi^M(Y)=Y,
		\end{equation}
		and describes a limit cycle of length $M$.
		Clearly, the \emph{fixed points} of $\Phi$, i.e.
		\begin{equation}
			 \Phi(Y)=Y 
		\end{equation}
		are (trivial) examples of limit cycles. On the other hand, the orbit of an eigenoperator $Y$ associated with an eigenvalue $\lambda$ which is not a root of unity does not close and is almost periodic.
		
		Importantly, according to Proposition 6.9 of~\cite{wolf2012quantum}, if $\mathcal{P}$ denotes the eigenprojection onto the fixed point space $\Fix(\Phi)$ of $\Phi$, then
		\begin{equation}
			X \in \Fix(\Phi) \Rightarrow \supp(X),\Ran(X) \subseteq \supp(\mathcal{P}(\mathbb{I}))\equiv \mathcal{H}_{0},
		\end{equation}
		where $\supp(A)$, $\Ran(A)$ stand for the support and the range of the operator $A$.
	Obviously, $\mathcal{P}(\mathbb{I})$ is a maximum-rank fixed point of $\Phi$. 
		
		Therefore, let us define $\tilde{\Phi}:\mathcal{B}(\mathcal{H}_{0})\mapsto \mathcal{B}(\mathcal{H}_{0})$ as
		\begin{equation}
			\tilde{\Phi}(X)=V^{\dagger} \Phi( VXV^{ \dagger } )V.
			\label{phit_def}
		\end{equation}
		Here $V:\mathcal{H}_{0} \mapsto \mathcal{H}$ is an isometry satisfying $V^{\dagger}V=\mathbb{I}_{\mathcal{H}_{0}}$ and $VV^{ \dagger }=Q$, the projection onto $\mathcal{H}_{0}$. $\tilde{\Phi}$ can be shown to be a faithful quantum channel (Lemma 6.4 of~\cite{wolf2012quantum}), namely having a full rank fixed point. We have
		\begin{equation}
			\Fix(\Phi)=0\oplus \Fix(\tilde{\Phi}),
		\end{equation}
		with the zero block acting on the orthogonal complement $\mathcal{H}_{0}^{\perp}$ of $\mathcal{H}_{0}$ and, analogously~\cite{albert2019asymptotics},
		\begin{equation}
			\Attr(\Phi)=0\oplus \Attr(\tilde{\Phi}) .
		\end{equation}
	\end{subsection}
	
	\begin{subsection}{Tomita-Takesaki modular theory}
		\label{tomita_intro}
		The modular theory by Tomita~\cite{Tom67a} and Takesaki~\cite{Takesaki1970,TakesakiII} was originally introduced to generalize integration and measure theory to non-Abelian algebras, but it turned out to be a fundamental and general tool in the algebraic description of quantum systems. For instance, it is used for $\sigma$-finite von Neumann algebras where it plays an important role in the classification of factors~\cite{Powers1967,connes_frencharticle,connes1995noncommutative}. Also, it is crucial in quantum statistical mechanics, as it associates with a faithful normal state a (modular) group of automorphisms over the algebra of observables. This establishes a connection with the Kubo-Martin-Schwinger (KMS) condition on equilibrium states~\cite{Kubo,MartinSchwinger,haagtd}. In recent years, left Hilbert algebras and modular theory were employed to obtain the algebra generated by a groupoid in the Schwinger's picture of quantum mechanics~\cite{groupoidI,groupoidII,feynmanngroupoid}. Incidentally, modular theory was also generalized to Jordan algebras, see~\cite{haagerup1984tomita}.
		
		We will give here a quick review of modular theory for finite-dimensional algebras.
		Let $\mathfrak M$ be a finite-dimensional $*$-algebra over a Hilbert space $\mathcal H$. By a well-known structure theorem~\cite[p.~74]{davidson1996c}, 
		the Hilbert space $\mathcal H$ can be decomposed as
		\begin{equation}
			\mathcal H=\mathcal H_0^\perp\oplus \bigoplus_{k=1}^M {\mathcal  H_{k,1}\otimes\mathcal H_{k,2}},
		\end{equation}
		so that the algebra $\mathfrak M$ decomposes as
		\begin{equation}
			\mathfrak M=0\oplus\bigoplus_{k=1}^M \mathcal M_{d_k}\otimes \mathbb I_{k,2}.
			\label{finite-algebra_dec}
		\end{equation}
		
		Here, $\mathcal H_0^\perp$ represents the degeneracy of the algebra, $\mathcal M_{d_k}$ is the algebra of matrices over $\mathcal H_{k,1}$, with $d_k=\dim \mathcal H_{k,1}$, $\mathbb{I}_{k,2}$ is the identity over the Hilbert space $\mathcal H_{k,2}$ and represents the multiplicity of the algebra $\mathcal{M}_{d_{k}}$~\cite{fagrkon}.
		By Riesz lemma, every state $\sigma$ over the algebra $\mathfrak M$~\cite{bratteli2012operator} can be represented as a density matrix belonging to the algebra:
		\begin{equation}\label{eq:faithfulstatetomitatakesaki}
			\sigma=\frac{1}{\sum_{k=1}^{M}\tr(\sigma_{k})} \left(0\oplus\bigoplus_{k=1}^{M}\sigma_k \otimes  \frac{\mathbb I_{k,2}}{m_k}\right)\in \mathfrak M,
		\end{equation}
		with $m_{k}=\dim \mathcal{H}_{k,2}$, so that $\sigma(A)=\braket{\sigma}{A}_{\mathrm {HS}}=\tr(\sigma A)$.	
		
		For faithful states $\sigma_k>0$, and an inner product can be defined on $\mathfrak M$ as
		\begin{equation}
			\braket{A}{B}_\sigma=\tr(\sigma A^\dagger B),
		\end{equation}
		which makes $(\mathfrak M,\braket{\ }{\ }_\sigma)$ a Hilbert space. 
		In particular, on this Hilbert space it is possible to obtain a cyclic representation of the algebra $\mathfrak M$~\cite{gelfand1943imbedding,segal,bratteli2012operator}.

		Starting from a faithful state $\sigma$, Tomita-Takesaki modular theory allows to obtain a group of unitary automorphisms on the algebra
		\begin{equation}\label{eq:tt_evolution_heisenberg_picture}
			A\mapsto\Delta^{-\mathrm i t} A\Delta^{\mathrm it},\quad t\in\mathbb R, \quad A\in\mathfrak M.
		\end{equation}
		The operator $\Delta:\mathfrak M\rightarrow \mathfrak M$ is self-adjoint with respect to the inner product $\braket{\ }{\ }_\sigma$, and it is obtained via the polar decomposition of the map $S:A\mapsto A^\dagger$. In the non-degenerate case ($\mathcal H_0^\perp=0$), the modular group takes the simple form
		\begin{equation}\label{eq:dynamicsexplicitfinitedimensional}
			A\mapsto \sigma^{-\mathrm it} A\sigma^{\mathrm it},
		\end{equation}
		with $\sigma>0$ (this can be easily generalized to the degenerate case by taking the maximum-rank state $\sigma=0\oplus \tilde\sigma$, with $\tilde\sigma>0$). Note that the unitary operator $\sigma^{\mathrm it}$ is in the algebra $\mathfrak M$.
		
		Conversely, starting from a one-parameter group of unitary automorphisms on the algebra $\mathfrak M$, one can obtain the equilibrium thermal state via a KMS condition~\cite{Kubo,MartinSchwinger}. In particular, the equilibrium thermal state corresponding to the dynamics~\eqref{eq:dynamicsexplicitfinitedimensional} depends on a parameter $\beta\in \mathbb R$ and it is in the form
		\begin{equation}
			\sigma_{\beta}=\frac{1}{Z}e^{-\beta H}=\frac{1}{\tr(\sigma^{\beta})}\sigma^{\beta}.
		\end{equation}
		Here, $H=-\log \sigma$ is the generator of the unitary evolution $\sigma^{\mathrm i t}$. Obviously,  the faithful state $\sigma$ in~\eqref{eq:faithfulstatetomitatakesaki} corresponds to the choice $\beta=1$.
	\end{subsection}
	\begin{subsection}{Structure theorem}
		The finer structure for $\Attr(\Phi)$ and the action of a quantum channel $\Phi$ on such subspace, i.e.\ of the peripheral map $\Phi_{P}=\Phi\vert_{\Attr(\Phi)}$ is given in the following theorem by Perez-Garcia and Wolf (Theorem 8 of~\cite{wolf2010inverse}).

		\begin{theorem}
			Let $\Phi$ be a quantum channel and $\mathcal{P}$ the eigenprojection onto its fixed point space $\Fix(\Phi)$.
			\begin{enumerate}
				\item  
				There exists a decomposition 
				\begin{equation}
					\mathbb{C}^{d}=\mathcal{H}_{0}^{ \perp} \oplus \bigoplus_{k=1}^{M} \mathcal{H}_{k,1}\otimes \mathcal{H}_{k,2},
					\label{H_dec}
				\end{equation}
				with $\mathcal{H}_{0}=\supp(\mathcal{P}(\mathbb{I}))$ and some Hilbert spaces $\mathcal{H}_{k,i}$ with $k=1,\dots,M$ and $i=1,2$, and positive definite density matrices $\rho_{k}$ on $\mathcal{H}_{k,2}$ such that
				\begin{equation}
					\Attr(\Phi)=0\oplus \bigoplus_{k=1}^{M}\mathcal{M}_{d_{k}} \otimes \rho_{k},
					\label{attr_struc}
				\end{equation}
				with $d_{k}=\dim(\mathcal{H}_{k,1})$, namely $X\in \Attr(\Phi)$ may be represented as
				\begin{equation}
					X=0\oplus \bigoplus_{k=1}^{M} x_{k} \otimes \rho_{k},
				\end{equation}
				for some matrices $x_{k}\in\mathcal{M}_{d_{k}}$;
				\item There exist unitary matrices $U_{k}$ on $\mathcal{H}_{k,1}$ and a permutation $\pi$ acting on 
				$\{ 1,\dots , M \}$ 				
				such that 
				\begin{equation}
					\Phi(X)=0 \oplus \bigoplus_{k=1}^{M} U_{k}x_{\pi(k)}U_{k}^{\dagger}\otimes \rho_{k},\; X \in \Attr(\Phi).
					\label{struc_phiP}
				\end{equation}
			\end{enumerate}
			\label{Wolf_struc}
		\end{theorem}

		\begin{remark}
			Each cycle of the permutation $\pi$ must act on matrices $x_{k}$ of the same order, consistently with the fact that $\Attr(\Phi)$ is an invariant subspace for $\Phi$ (Proposition 6.12 of~\cite{wolf2012quantum}), i.e.
			\begin{equation}
				\Phi(\Attr(\Phi))=\Attr(\Phi).
				\label{attr_inva}
			\end{equation}  
		\end{remark}
		
		\begin{remark}
		\label{mod_prod}
			In the faithful case we have that $\dim(\mathcal{H}_{0}^{ \perp})=0$ and the zero factor in~\eqref{attr_struc} disappears. Furthermore, 
					it turns out that $\Attr(\Phi^{\dagger})$ is a unital $\ast$-algebra (Theorem 1 of~\cite{carbone2020period}) with the following structure~\cite{davidson1996c}
			\begin{equation}
				\Attr(\Phi^{\dagger})=\bigoplus_{k=1}^{M} \mathcal{M}_{d_{k}} \otimes \mathbb{I}_{k,2},
			\end{equation}
			i.e.~\eqref{finite-algebra_dec} without the zero first block.
			In the language of~\cite{Viola_inform_pres}, $\Attr(\Phi)$ is a distorted algebra, %of $\Attr(\Phi^{\dagger})$ 
			in the sense that it is an algebra under the star product
			\begin{equation}
				A \star B := A\mathcal{P}_{P}(\mathbb{I})^{-1}B \in \Attr(\Phi),\quad A,B\in \Attr(\Phi),
				\end{equation}
where $\mathcal{P}_{P}$ denotes the spectral projection onto $\Attr(\Phi)$. See also~\cite{contraction,contraction2} for the contracted algebra of a dissipative quantum system.
			\end{remark}
		
		\begin{remark} 
			More generally, Theorem~\ref{Wolf_struc} is valid for Schwarz maps~\cite{wolf2010inverse,wolf2012quantum}, namely positive trace-preserving maps obeying to the operator inequality
			\begin{equation}
				\Phi^{\dagger}(A^{\dagger})\Phi^{\dagger}(A) \leqslant \Phi^{\dagger}(A^{\dagger}A),\quad A \in \mathcal{B}(\mathcal{H}).
			\end{equation}
		\end{remark}
	\end{subsection}
	
\end{section}
\begin{section}{Unitary asymptotic dynamics}
	\label{unit_per}
	One of the most intriguing features of Theorem~\ref{Wolf_struc} is the occurrence of permutations in the asymptotic action~\eqref{struc_phiP} of $\Phi$, which makes the peripheral map $\Phi_{P}$, i.e.\ the asymptotic dynamics, not generally unitary. Thus one may ask about the conditions on $\Phi$ under which the map $\Phi_{P}$ is unitary. Two sufficient conditions are provided in the following Theorem~\cite{AFK_asympt}.
	
	\begin{theorem}
		Given a quantum channel $\Phi$ with corresponding peripheral map $\Phi_{P}$, then
		\begin{enumerate}
			\item If $\Phi=e^{\mathcal{L}}$, with $\mathcal{L}$ being a GKLS generator~\cite{GKS_76,Lindblad_76} (Markovian channel), then $\Phi_{P}(X)=U X U^{\dagger}$, with $U$ unitary,
			\item If $\Phi^{2}=\Phi$ (idempotent channel), then $\Phi_{P}(X)=U X U^{\dagger}$, with $U$ unitary.
		\end{enumerate} 
		\label{suff_unit}
	\end{theorem}
	The proof, which may be found in \cite{AFK_asympt}, is based on  a characterization of the absence of permutations in the structure of the asymptotic map.

	\begin{remark}
		The first sufficient condition is in line with the asymptotics of quantum dynamical semigroups~\cite{albert2014symmetries}.
	\end{remark}
	\begin{remark}
		Note that both conditions are not necessary, since a Markovian channel is invertible (as a linear map), while an idempotent channel, different from the identity, is not.
	\end{remark}
	
	Trivially, the peripheral channel $\Phi_P$ is unitary when there are no permutations in the action~\eqref{struc_phiP}. Indeed, as we will show in Section~\ref{Wolf-TT}, in such a case the dynamics is obtained from the modular group corresponding to a faithful fixed state.
	
	Now, we will provide a characterization for the unitarity of the asymptotic dynamics $\Phi_P$.
	\begin{theorem}
	Let $\Phi$ be a quantum channel with attractor subspace $\Attr(\Phi)$ of the form~\eqref{attr_struc} and peripheral map $\Phi_P$. 

Then 
\begin{equation}
	\Phi_P(X)=U X U^\dagger,  \qquad \text{with}\quad X\in \mathrm{Attr}(\Phi),
\end{equation}
 for some unitary operator $U$ on $\mathcal{H}$, iff $\rho_k\sim\rho_{\pi(k)}$, namely 
\begin{equation}
\label{ch_unit_as-map}
	\rho_{k}=V_k \rho_{\pi(k)} V_k^\dagger,
\end{equation}
 for some unitaries $V_k:\mathcal H_{\pi(k),2}\rightarrow \mathcal H_{k,2}$.
\label{ch_unit_as_dyn}
	\end{theorem}
\begin{proof}
	Besides being necessary, one can easily show that this condition is also sufficient. Indeed, if $\Phi$ is faithful, then we can define the operator $U:\mathcal H_0\rightarrow \mathcal H_0$ as
	\begin{equation}
		U: \bigoplus_{k=1}^{M} \phi_{k} \otimes \psi_{k}\mapsto \bigoplus_{k=1}^{M} U_{k}\phi_{\pi(k)} \otimes V_{k}\psi_{\pi(k)},
	\end{equation}
	and its extension to $\mathcal H_0$ follows by linearity.
	It is easy to see that this is indeed unitary on $\mathcal H_0$, and it can be extended on the whole $\mathcal H$ as $\mathbb I_0\oplus U$. Moreover, it satisfies the condition stated in Theorem \ref{ch_unit_as_dyn}.
	\end{proof}

Observe that condition~\eqref{ch_unit_as-map} holds for unital quantum channels, i.e.\ $\Phi(\mathbb I)=\mathbb I$. Indeed, a unital quantum channel satisfies conditions $\rho_k=\mathbb I_{k,2}/m_k$ and $m_k=m_{\pi(k)}$ for all $k=1, \dots , M$. Moreover, unitarity also ensures Hilbert-Schmidt unitarity, that is
	\begin{equation}
		\norm{\Phi(X)}_{\mathrm{HS}}=\norm{X}_{\mathrm{HS}},\quad X\in\mathrm{Attr}(\Phi).
	\end{equation}
	 As a final remark, it is possible to show~\cite{AFK_asympt} that every map defined on a set in the form~\eqref{attr_struc}, and acting as~\eqref{struc_phiP} can be extended to a quantum channel over $\mathcal B(\mathcal H)$. This can be interpreted as the inverse of structure theorem~\ref{Wolf_struc}.
\end{section}
\begin{section}{Asymptotic dynamics and modular theory}
	\label{Wolf-TT}
	In this Section, we are going to study the relation between the asymptotic dynamics of a quantum channel and  Tomita-Takesaki modular theory. In particular, we want to find the conditions under which the asymptotic dynamics $\Phi_{P}$ can be obtained from the modular evolution group~\eqref{eq:dynamicsexplicitfinitedimensional} associated with an equilibrium state. In doing this, we will use the structure theorem~\ref{Wolf_struc}, and we will write the modular group corresponding to fixed states.
	
	Let us assume that the quantum channel $\Phi$ is faithful. We could drop this requirement simply by considering the reduced channel $\tilde\Phi$ defined in equation~\eqref{phit_def}.
	
	We start by considering a particular situation, in which the permutation in equation~\eqref{struc_phiP} has only one cycle, i.e., up to a relabeling, it is in the form
	\begin{equation}
		\pi(j)=j+1\quad \mathrm{mod}\ M.
	\end{equation}
	A state $\sigma$ in the attractor space $\Attr(\Phi)$ in~\eqref{attr_struc} will be a fixed point whenever its components $\sigma_j$ satisfy the equation 
	\begin{equation}
		\sigma_j=U_j \sigma_{j+1} U_j^\dagger\quad \mathrm{mod}\ M, \quad 1\leqslant j\leqslant M.
		\label{eq:fixed_points_equation}
	\end{equation}
	By setting $V_1=U_1U_2\dots U_M$, equation~\eqref{eq:fixed_points_equation} reads
	\begin{align}
		\sigma_1&=V_1\sigma_1 V_1^\dagger,\nonumber\\
		\sigma_{j}&=U_{j-1}^\dagger \cdots U_1^\dagger\sigma_1 U_1\cdots U_{j-1},\quad 2\leqslant j\leqslant M,
	\end{align}
	so that the fixed points have $\sigma_1$ in the commutant 
	\begin{equation}
		\{V_1\}'=\{ A \in \mathcal{M}_{d_{1}} \,\vert\, [A,V_1]=0 \},
	\end{equation}	
	of $V_1$, where $[\cdot,\cdot]$ stands for the commutator.
	Writing the spectral decomposition of $V_1$
	\begin{equation}
		V_1=\sum_{j=1}^{n} e^{i\theta_j}P_j^{(1)},\quad \theta_j\in \mathbb{R},
	\end{equation}
	we have $\sigma_1=\sum_{j=1}^{n} P_j^{(1)}\sigma_1P_j^{(1)}$. We make another requirement, namely $\sigma_1$ being in the bicommutant $\{V_1\}''\subset\{V_1\}'$. In this way, it will be a function of $V_1$ in the form
	\begin{equation}
		\sigma_1=\sum_{j=1}^n \lambda_{j} P_j^{(1)}.
	\end{equation}
	The other components $\sigma_k$ will have a similar structure
	\begin{equation}
		\sigma_k= U_{k-1}^\dagger \cdots U_1^\dagger \sigma_1 U_1\cdots U_{k-1}=\sum_{j=1}^{n} \lambda_j P_j^{(k)},
	\end{equation}
	with $P_j^{(k)}=U_{k-1}^\dagger \cdots U_1^\dagger P_j^{(1)} U_1\cdots U_{k-1}$. 
	
	Note that we can also apply Tomita-Takesaki modular theory to the space $\Attr(\Phi)$, being it a von Neumann algebra with respect to a modified product (see Remark~\ref{mod_prod}). In the Schr\"odinger picture, the unitary evolution~\eqref{eq:dynamicsexplicitfinitedimensional} generated by $\sigma$ via modular theory is
	\begin{equation}\label{eq:tomita_takesaki_schrodinger}
		X\mapsto \sigma^{\mathrm i t}X \sigma^{-\mathrm it}
	\end{equation}
	with
	\begin{equation}
		\sigma^{\mathrm it}=\bigoplus_{k=1}^M\sigma_k^{\mathrm it}\otimes \rho_k^{\mathrm it}=\bigoplus_{k=1}^M\left(\sum _{j=1}^{n} e^{\mathrm i t\log \lambda _j}P_j^{(k)}\right)\otimes \rho_k^{\mathrm it}.
	\end{equation}

	We see that the permutation disappears, and the evolution turns out to be unitary. By setting $\lambda_j=e^{\theta_j}$, we see that the evolution for $t=1$ is just
	\begin{equation}
		\sigma^{\mathrm  i}X\sigma^{-\mathrm i}=\Phi_V(X),\quad X\in \mathrm{Attr}(\Phi) 
	\end{equation}
	where $\Phi_V$ is the map
	\begin{equation}
		\Phi_V(X)= \bigoplus_{k=1}^M V_k X_k V_k^\dagger \otimes \rho_k,\quad X\in \mathrm{Attr}(\Phi)
	\end{equation}
	with
	\begin{equation}
		V_k=U_k U_{k+1}\cdots U_M U_1\cdots U_{k-1}=U_{k-1}^\dagger \cdots U_1^\dagger V_1 U_1\cdots U_{k-1}.
	\end{equation}
	
	Note that the map $\Phi_V$ is just $\Phi_P^M$, the peripheral evolution repeated over a complete cycle. The generalization to a permutation with more cycles can be obtained in a similar way. Indeed, the corresponding Tomita-Takesaki evolution turns out to be the $M$-th power of the asymptotic map $\Phi_{P}$, with $M$ being the least common multiple of the lengths of the cycles of the permutation. In particular, this corresponds to the asymptotic evolution if and only if there are no permutations.
	
\end{section}

\begin{section}{Conclusions and outlooks}
	In this Article, we gave further insights into the asymptotics of finite-dimensional open quantum systems, whose study has been strongly motivated by quantum information protection and processing~\cite{zanardi1997noiseless,zanardi2014coherent,Viola_inform_pres}. Theorem~\ref{Wolf_struc}, giving the structure of the attractor subspace and the action of the channel on it, reveals two important facts. 
	
	First, up to a zero block in the non-faithful case, the attractor subspace is a $*$-algebra with respect to the modified product~\eqref{mod_prod}. 
	
	Second, in the structure of the asymptotic map, partial permutations between the factors in the decomposition~\eqref{attr_struc} may occur according to~\eqref{struc_phiP}, making the dynamics at large times no longer unitary. 
	
	The occurrence of permutations also explains how the asymptotic evolution cannot always be regarded as the modular dynamics on the attractor subspace associated with a fixed state, as shown in Section~\ref{Wolf-TT}. More precisely,  Tomita-Takesaki evolution is the (unitary) peripheral map of the $M$-times iterated dynamics, with $M$ denoting the least common multiple of the lengths of the cycles of the permutation. This unveils a connection between permutations and the divisibility of the quantum channel which generates the dynamics under consideration, deserving further studies in the future~\cite{AFK_asympt}.
	
	Also, as already illustrated in the faithful case by Novotn{\'y}, Mary{\v s}ka, and Jex~\cite{jex_st_2018}, the asymptotic evolution may not be a Hilbert-Schmidt unitary but it is still unitary with respect to a modified scalar product. In other words, the peripheral map is not generally unitary  because of permutations, but it can be shown to be unitary in a weaker sense.
	
	Another open problem could be the connection between eventually and asymptotically entanglement breaking maps~\cite{paulsen2018asymptotically,rahaman2018eventually} and the (non-)unitarity of the asymptotic map~\cite{hanson2020eventually}. Entanglement breaking maps~\cite{horodecki2003entanglement} have turned out to be a central topic in the theory of quantum information, specifically in entanglement transferring~\cite{amendingdepasquale,depasqualecut}. In particular, many efforts have been devoted to the proof of the PPT square conjecture~\cite{christandl2012ppt}, recently achieved in the finite-dimensional case~\cite{majewski2021ppt}.
\end{section}
\begin{section}{Acknowledgements}
This work was partially supported by Istituto Nazionale di Fisica Nucleare (INFN) through the project ``QUANTUM'', by Regione Puglia and QuantERA ERA-NET Cofund in Quantum Technologies (Grant No. 731473), project PACE-IN, and by the Italian National Group of Mathematical Physics (GNFM-INdAM).
\end{section}

\bibliography{mybibfile.bib}
\bibliographystyle{spmpsci}

\end{document}